\documentclass[12pt, prd, showpacs]{revtex4}
\usepackage{amssymb}
\usepackage{amsmath}

\input{tcilatex}

\begin{document}

\title{Acceleration of particles by black holes - general explanation}
\author{Oleg B. Zaslavskii}
\affiliation{Department of Physics and Technology, Kharkov V.N. Karazin National
University, 4 Svoboda Square, Kharkov, 61077, Ukraine}
\email{zaslav@ukr.net}

\begin{abstract}
We give simple and general explanation to the effect of unbound acceleration
of particles by black holes. It is related to the fact that the scalar
product of a timelike vector of the four-velocity of an ingoing particle and
the lightlike horizon generator tends to zero in some special cases, so the
condition of "motion forward in time" is marginally satisfied. In this
sense, an ingoing particle with special relation between parameters imitates
the property of infinite redshift typical of any outgoing particle near the
future horizon of \ a black hole. We check this assertion \ using the
Reissner-Nordstr\"{o}m and rotating axially-symmetric metrics as examples.
\end{abstract}

\keywords{black hole horizon, acceleration, light cone}
\pacs{04.70.Bw, 97.60.Lf, 04.40.Nr }
\maketitle




\section{Introduction}

Recently, an interesting observation was made in \cite{ban} about
acceleration of particles near the horizon of a rotating black hole to
unlimited energies $E_{c.m.}$ in the centre of mass frame. In this sense, a
black hole can act a cosmic supercollider, so in its vicinity new physics is
expected at the Planck scale. A series of papers followed where details of
this process were studied \cite{gp4} - \cite{bsw} and its generalization 
\cite{prd} and extension to charged nonrotating black holes \cite{jl} were
suggested. The goal of the present work is to give a general and
comprehensive explanation to this interesting effect. Rather surprisingly,
it turns out that such an explanation is very simple and relies not on the
details of theory but on the mutual properties of particles and a light cone
near the future horizon of \ a black hole. Thus, we generalize previous
observations and elucidate the underlying reason for the effect for the
variety of metrics previously considered.

\section{Basic formulas}

It would seem that the effect connected with acceleration of particles
requires necessarily detailed analysis of their equations of motion. It is
just the approach developed in previous works \cite{ban} - \cite{jl}.
Instead, in the present work we focus attention on what happens to the
four-velocity of a particle with respect to its local light cone in the
immediate vicinity of the horizon. Let us consider the collision of two
particles near the future horizon of \ a black hole. In doing so, one should
clearly distinct two different cases: 1) particles move in the opposite
directions (towards the horizon and away from it), 2) both particles move
towards the horizon. Actually, the first case was discussed in \cite{pir}
(although the corresponding condition was not explicitly pronounced there) a
long time ago. The second case is discussed in the series of aforementioned
papers \cite{ban} \ - \cite{jl}.

We will use the following geometric construction. Let us introduce in the
point $P$ under consideration and its vicinity the tetrad with lightlike
vectors $l^{\mu }$, $N^{\mu }$ and spacelike vectors $a^{\mu }$, $b^{\mu }$
orthogonal to them. Here, the vectors $l^{\mu }$, $N^{\mu }$ are normalized,
say, as $l^{\mu }N_{\mu }=-1$.

Then,%
\begin{equation}
g_{\alpha \beta }=-l_{\alpha }N_{\beta }-l_{\beta }N_{\alpha }+\sigma
_{\alpha \beta }  \label{met}
\end{equation}%
where $\sigma _{\alpha \beta }=a_{\alpha }b_{\beta }+a_{\beta }b_{\alpha }$, 
$l^{\alpha }\sigma _{\alpha \beta }=N^{\alpha }\sigma _{\alpha \beta }=0$
(see, for example, textbook \cite{erik}). We assume that it is the vector $%
l^{\mu }$ that becomes the generator of the future horizon. In general, we
can use the decomposition of the four-velocity $u^{\mu }$ in the form%
\begin{equation}
u_{i}^{\mu }=\frac{l^{\mu }}{2\alpha _{i}}+\beta _{i}N^{\mu }+s_{i}^{\mu }%
\text{, }s_{i}^{\mu }=A_{i}a^{\mu }+B_{i}b^{\mu }  \label{u}
\end{equation}%
where $i=1,2$ labels the particles and $\alpha _{i}$, $\beta _{i}$, $A_{i}$
and $B_{i}$ are coefficients. The time-like vector $u^{\mu }$ is normalized
as usual, $(uu)=-1$, hereafter the symbol $(...)$ denotes the scalar
product. Then, it follows from (\ref{u}) that%
\begin{equation}
\beta _{i}=-(u_{i}l)\text{, }  \label{b}
\end{equation}%
\begin{equation}
\alpha _{i}=-\frac{1}{2}(u_{i}N)^{-1}\text{.}  \label{alpha}
\end{equation}%
As vectors $u^{\mu }$, $l^{\mu }$, $N^{\mu }$ are assumed to be
future-directed, $\alpha _{i}>0$, $\beta _{i}>0$ (motion "forward in time").
The normalization condition entails%
\begin{equation}
s_{i}^{\mu }s_{i\mu }=\frac{\beta _{i}}{\alpha _{i}}-1\text{.}  \label{s}
\end{equation}%
The case $\beta _{i}=\alpha _{i}$, $s_{i}^{\mu }=0$ corresponds to pure
radial motion (see below).

Then,%
\begin{equation}
-(u_{1}u_{2})=\frac{1}{2}(\frac{\beta _{1}}{\alpha _{2}}+\frac{\beta _{2}}{%
\alpha _{1}})-(s_{1}s_{2})\text{.}
\end{equation}%
The energy in the centre of mass frame \cite{ban} - \cite{jl} is equal to $%
E_{c.m.}^{2}=m_{1}^{2}+m_{2}^{2}-2m_{1}m_{2}(u_{1}u_{2})$ ($m_{i}$ are rest
masses of particles), so%
\begin{equation}
E_{c.m.}^{2}=m_{1}^{2}+m_{2}^{2}+m_{1}m_{2}[\frac{\beta _{1}}{\alpha _{2}}+%
\frac{\beta _{2}}{\alpha _{1}}-2(s_{1}s_{2})]\text{.}  \label{en}
\end{equation}%
.

\section{Ingoing versus outgoing particles in the vicinity of the horizon:
general approach}

\subsection{Case 1.}

Let particle 1 be going from the immediate vicinity of the horizon in the
outward direction. We are dealing with the future horizon of a black hole,
the vector $l^{\mu }$ becoming its generator when the horizon is approached.
Meanwhile, this particle, by assumption, cannot cross the horizon and does
not penetrate the region inside. Therefore, near the horizon it does not
move in the direction of $N^{\mu }$, so it moves almost in the direction of
the horizon generator $l^{\mu }$. Hence, the component of the four-velocity
in the direction of $N^{\mu }$ should vanish, so it follows from (\ref{u})
that

\begin{equation}
\beta _{1}\rightarrow 0\text{.}  \label{con}
\end{equation}%
Now, it is worth noting that by construction, the vector $s^{\mu }\,$\ is
spacelike, so $(ss)>0$ for $s^{\mu }\neq 0$. Then, it follows from (\ref{s})
that (\ref{con}) entails also 
\begin{equation}
\alpha _{1}\rightarrow 0\text{.}  \label{con2}
\end{equation}%
Meanwhile, $\alpha _{2}$ is arbitrary positive quantity. Then, it is seen
from (\ref{en}) that $E_{c.m.}^{2}\rightarrow \infty .$ One can say that
this is just direct consequence of infinite redshift near the horizon. In
the examples below it is checked that eq. (\ref{con}) is indeed satisfied.

\subsection{Case 2}

This case (both particles move towards the horizon) is much more interesting
since the frame of the centre of mass falls down with both particles \cite%
{ban}, so the possible effect of unbound acceleration is not direct
manifestation of the redshift. In general, as it is seen from (\ref{en}), $%
E_{c.m.}^{2}$ remains finite even in the vicinity of the horizon for any
nonzero $\alpha _{1}$, $\alpha _{2}$. Basically, the simple point here is
that, for unbounded collision energies to occur in this case, certain
conditions on the parameters of the particle need to be satisfied (as
discovered by previous works), and these conditions are equivalent to the
requiring that $\alpha _{1}$ vanish as the particle approaches the horizon.
Indeed, let us now \textit{assume }that (\ref{con}) and, hence, (\ref{con2})
hold now (in case 1 they were satisfied automatically). In other words, an
ingoing particle imitates the property of infinite redshift (\ref{con}), (%
\ref{con2}) typical of an outgoing particle near the horizon. Then, again it
follows from (\ref{en}), (\ref{con}), (\ref{con2}) that $E_{c.m.}^{2}%
\rightarrow \infty .$ This is just the effect discovered in \cite{ban} and
studied in \cite{gp4} - \cite{jl}. Thus, in case 2 the special condition (%
\ref{con}) is needed. It relates the parameters of a particle like the
energy and angular momentum or the energy and electric charge, etc. (see
examples below).

The above observations can be also reformulated as follows. Consider the
vector $\xi ^{\mu }$ which is timelike in the region where particles
approach the horizon, $N^{2}=-(\xi \xi )>0$: 
\begin{equation}
\xi ^{\mu }=\frac{1}{2}l^{\mu }+N^{2}N^{\mu }\text{.}  \label{ksi}
\end{equation}

We can easily deduce two additional properties.

1) Let, in the near-horizon limit, condition (\ref{con}) for some particle
be satisfied, and let $(\xi u)$ be finite (otherwise the particle is
arbitrary). Then, the vector $\xi ^{\mu }$ becomes lightlike in this limit.

\textit{Proof. }It follows from\textit{\ }(\ref{u}), (\ref{con}) - (\ref{ksi}%
)\textit{\ }that\textit{\ }in this limit $(\xi u)\approx N^{2}(Nu)=-\frac{%
N^{2}}{2\alpha }$. As this quantity is finite, it follows from (\ref{con})
that also $N\rightarrow 0$.

2) Let us, instead of (\ref{con}), assume that $(\xi u)$ $\rightarrow 0$.
Then, (\ref{con}) is satisfied and the vector $\xi ^{\mu }$ becomes
lightlike in this limit.

\textit{Proof.} Multiplying (\ref{ksi}) by $u_{\mu }$, we observe that both
terms are negative. Therefore, each of them vanishes separately in this
limit, so $\alpha \rightarrow 0$, $N^{2}\rightarrow 0$. As a consequence, $%
E_{c.m.}^{2}\rightarrow \infty $.

The situation where the vector $\xi ^{\mu }$ is timelike in some region but
becomes lightlike on some hypersurface is typical of Killing horizons.
However, we would like to emphasize that nowhere did we use Killing
equations. It is worth also noting that in the formulation of statements 1)
and 2) we relied on one particle with the four-velocity $u^{\mu }$, so these
statements are not related to the collision of two particles directly.

The results under discussion can be reexpressed in another way with the help
of Kruskal-like coordinates. Let, for simplicity, the metric be written in
the form%
\begin{equation}
ds^{2}=-CdUdV+\gamma _{ab}dx^{a}dx^{b}
\end{equation}%
where $a=1,2$ and the metric coefficients are regular functions of the
coordinates $U$ and $V$ (this is certainly possible for the nonrotating
black holes). Here, the coordinates, $x^{a}$ have the meaning of angular
coordinates in the spherically symmetric case. On the horizon $U=0$ or $V=0$%
. Then, repeating the above arguments, we see that it follows from (\ref{con}%
) that, say, near the horizon $U=0$ the component of the four-velocity $%
u^{U}\sim \beta \rightarrow 0$. Taking into account the regularity of the
metric, we can write that $\alpha \sim \beta \sim U$, whence we have 
\begin{equation}
\frac{dU}{d\tau }\sim U\text{,}
\end{equation}%
so%
\begin{equation}
\tau \sim -\ln U\rightarrow \infty
\end{equation}%
in accordance with previous results for the Kerr \cite{gp4}, \cite{gpm} or
Reissner-Nordstr\"{o}m \cite{jl} black holes.

Let us now illustrate these general properties by two examples.

\section{Examples}

\subsection{Radial motion in Reissner-Nordstr\"{o}m black hole}

\begin{equation}
ds^{2}=-dt^{2}N^{2}+\frac{dr^{2}}{N^{2}}+r^{2}d\omega ^{2}.
\end{equation}

Equivalently, the metric can be rewritten in the form

\begin{equation}
ds^{2}=-dt^{2}N^{2}+dn^{2}+r^{2}d\omega ^{2}.
\end{equation}%
Here $n$ has the meaning of the proper distance, $d\omega ^{2}=\sin
^{2}\theta d\phi ^{2}+d\theta ^{2}$, $N^{2}=1-\frac{2M}{r}+\frac{Q^{2}}{r^{2}%
}$ where $M$ is the black hole mass, $Q$ is its charge. The event horizon
lies at $\ r=r_{H}=M+\sqrt{M^{2}-Q^{2}}$. \ Consider radial motion of a
particle having the charge $q$ and rest mass $m$. From the equations of
motion one finds the components of the four-velocity for a pure radial
motion:%
\begin{equation}
u^{0}=\frac{X}{N^{2}m}\text{, }u^{1}=\varepsilon \frac{Z}{mN}  \label{u1}
\end{equation}%
where $\varepsilon =-1$ for the direction towards the horizon and $%
\varepsilon =+1$ for the opposite direction of motion,%
\begin{equation}
X=E-\frac{qQ}{r},Z=\sqrt{X^{2}-m^{2}N^{2}},  \label{xz}
\end{equation}%
the coordinates are $x^{0}=t,x^{1}=n$, $x^{2}=\theta $, $x^{3}=\phi $.

Here, $E$ is the conserved energy, dot denotes differentiation with respect
to the proper time $\tau $, $u^{\mu }$ is the four-velocity. The quantity $%
X_{H}=E-\frac{qQ}{r_{H}}\geq 0$, so it is positive for all $r>r_{H}$ (motion
"forward in time"). Then, the vector (\ref{ksi}) has the components $\xi
^{\mu }=(1,0,0,0)$ and coincides with the Killing vector. Let us also
introduce two lightlike vectors 
\begin{equation}
l^{\mu }=(1,N,0,0)\text{, }N^{\mu }=\frac{1}{2}(\frac{1}{N^{2}},-\frac{1}{N}%
,0,0),
\end{equation}%
$(Nl)=-1$. The vectors $a_{\mu }$ and $b_{\mu }$ have nonzero components $%
a_{\theta }=r$, $b_{\phi }=r\sin \theta $. One can check that the equality (%
\ref{met}) is satisfied.

Then,%
\begin{equation}
-(\xi u)=\frac{X}{m}  \label{al}
\end{equation}%
and, according to (\ref{b}),%
\begin{equation}
\beta =-(ul)=\frac{X-\varepsilon Z}{m}>0\text{. }  \label{ul}
\end{equation}%
The quantity $-(uN)N^{2}=\frac{1}{2}\frac{X+\varepsilon Z}{m}>0$ is finite
for both signs of $\varepsilon $ in agreement with discussion in Sec. II, so
it follows from (\ref{alpha}) that%
\begin{equation}
\alpha =\frac{mN^{2}}{X+\varepsilon Z}\text{.}  \label{un}
\end{equation}%
Bearing in mind that $X^{2}-Z^{2}=m^{2}N^{2}$, it is easy to see that 
\begin{equation}
\beta =\alpha  \label{eq}
\end{equation}%
in accordance with what is said about pure radial motion in Sec. II.

\subsubsection{Case 1}

Let us take $\varepsilon =+1$ in the expression (\ref{u1}) for $u^{1}$ that
corresponds to the motion away from the horizon towards infinity. Then, it
follows from (\ref{ul}), (\ref{eq}) that 
\begin{equation}
\alpha =\frac{X-Z}{m}\text{.}  \label{a+}
\end{equation}%
Outside the horizon, $\alpha >0$. In the horizon limit $N\rightarrow 0$ and
it is seen from (\ref{xz}) that $Z\rightarrow X$ in this limit, so for any
particle irrespective of the relation between the parameters $\alpha
\rightarrow 0$ in accordance with general discussion of case 1 in Sec. III A

\subsubsection{Case 2.}

Now, $\varepsilon =-1$. On the horizon $Z=X_{H}$ (hereafter we use subscript
"H" for the values calculated on the horizon), $\alpha _{H}=\frac{2X_{H}}{m}%
\geq 0$, $-(\xi u)_{H}=\frac{X_{H}}{m}\geq 0$. If for particle 1 $X_{H}=0$, $%
qQ=Er_{H}$, it follows that $\alpha _{1}=\beta _{1}\rightarrow 0$, when the
horizon is approached. Then, the above consideration applies which leads to
the result $E_{c.m.}^{2}\rightarrow \infty $. that agrees with the one
obtained earlier \cite{jl}.

\subsection{Axially-symmetric rotating black hole}

Now, let us consider the generic metric describing an axially-symmetric
black hole%
\begin{equation}
ds^{2}=-N^{2}dt^{2}+g_{\phi \phi }(d\phi -\omega dt)^{2}+dl^{2}+g_{zz}dz^{2}
\label{z}
\end{equation}

that includes the Kerr and Kerr-Newman black holes. However, the
configuration is more general due to the possible presence of matter (dirty
black holes). We want to compare general formalism of Sections II and III
with the more standard approach based on equations of motions. For metric (%
\ref{z}), it follows from equations of motion that

\begin{equation}
\dot{t}=u^{0}=\frac{X}{N^{2}},\text{ }X=E-\omega L  \label{t}
\end{equation}%
(for simplicity, here we assume that the rest mass $m=1$). 
\begin{equation}
\dot{\phi}=\frac{L}{g_{\phi \phi }}+\frac{\omega X}{N^{2}},  \label{phi}
\end{equation}%
\begin{equation}
\dot{l}=\varepsilon \frac{Z}{N}\text{, }Z^{2}=X^{2}-N^{2}(1+\frac{L^{2}}{%
g_{\phi \phi }})  \label{n}
\end{equation}%
where $u_{0}=-E$ is the energy, $u_{\phi }=L$ is the angular momentum, $%
\varepsilon =\pm 1$ has the same meaning as before. For motion "forward in
time", we must have $\dot{t}>0$, so $E-\omega L>0$.

Now, the relevant lightlike vectors are%
\begin{equation}
l_{\mu }=(-N^{2},N,0,0)
\end{equation}%
\begin{equation}
N_{\mu }=\frac{1}{2N^{2}}(-N^{2},-N,0,0)
\end{equation}%
\begin{equation}
(Nl)=-1\text{.}
\end{equation}%
The vector (\ref{ksi}) reads%
\begin{equation}
\xi ^{\mu }=\xi _{1}^{\mu }+\omega \xi _{2}^{\mu }
\end{equation}%
where $\xi _{1}^{\mu }=(1,0,0,0)$ is the Killing vector that generates
translations in time, $\xi _{2}^{\mu }=(0,0,1,0)$ generates rotations. On
the horizon $N=0$ the vector $\xi ^{\mu }$ becomes lightlike.

One can check that eq. (\ref{met}) is indeed satisfied, where nonzero
components of vectors $a_{\mu }$ and $b_{\mu }$ equal $b_{z}=\sqrt{g_{zz}}$%
., $a_{\phi }=\sqrt{g_{\phi \phi }}$, $a_{0}=-\omega a_{\phi }$, The scalar
product $(ua)=\frac{L}{\sqrt{g_{\phi \phi }}}$ is finite. Then, similar to
what we had in the Reissner-Nordstrom case, one finds that $-(u\xi )=X$ and
eqs. (\ref{ul}), (\ref{un}) hold where now $Z$ is defined in (\ref{n}).
Then, one can easily obtains that in this case,%
\begin{equation}
\frac{\beta }{\alpha }=1+\frac{L^{2}}{g_{\phi \phi }}\text{.}  \label{bal}
\end{equation}%
instead of (\ref{eq}). Although $\beta $ and $\alpha $ are not equal now,
they are proportional to each other, so if $\beta \rightarrow 0$, also $%
\alpha \rightarrow 0$ in accordance with general discussion in Sec. III A.

\subsubsection{Case 1}

For the motion away from the horizon, $\varepsilon =1$. In the horizon limit 
$N\rightarrow 0$, one can see from (\ref{n}) that $Z\rightarrow X$, so we
again obtain the properties (\ref{con}), (\ref{con2}) for any relationship
between the energy and the angular momentum of a particle.

\subsubsection{Case 2}

Now, in the horizon limit $\beta \rightarrow 2X_{H}$. The critical value is
singled out by the condition $X=0$ on the horizon ($E=\omega _{H}L$) that
indeed coincides with (\ref{con}). Then, we again obtain that q$%
E_{c.m.}^{2}\rightarrow \infty $ in accordance with the previous discussion
and \cite{prd}.

For completeness, we should make a reservation. Apart from the cases $\dot{l}%
>0$ ($\varepsilon =+1)$ and $\dot{l}<0$ ($\varepsilon =-1$) in both examples
there exist special orbits for which $\dot{l}=0$ (see \cite{bar} for the
discussion of these orbits in the case of the Kerr metric). It is seen from (%
\ref{u1}), (\ref{xz}) or (\ref{n}) that for such orbits $Z=0$, $X\sim N$. In
the horizon limit, $N\rightarrow 0$, $X\rightarrow 0$ and we again return to
the condition (\ref{con}).

\section{Conclusions}

Thus, we elucidated the generic nature of the effect and showed that
diversity of different metrics and even classes of metric has the same
underlying reason in this context. In doing so, we did not use explicitly
the equations of motion of particles at all, did not rely on an explicit
form of the metric, field equations from which it is obtained, etc. (We used
equations of motions to compare two different approaches only.) Actually,
the nature of the effect turned out to be surprisingly simple and stemming
from the mutual properties of lightlike and timelike vectors in the vicinity
of the future horizon. It may happen that the condition (\ref{met}) is not
realized in some particular cases (say, for some classes of trajectories 
\cite{lake2}). Nonetheless, if (i) the horizon exists and (ii) the condition
(\ref{con}) is indeed satisfied, the effect of unbound $E_{c.m.}$ can
manifest itself in general. Moreover, it follows from our derivation that
these reasonings apply not only to the horizons of static or stationary
black holes. As a matter of fact, the effect is valid even if the
aforementioned condition is obeyed for some portion of the surface only.
Moreover, these portions can shrink to the point. In particular, the results
of the present work seem to apply to dynamic or isolated horizons \cite{ash}.

The fact that the essence of the effect of infinite \thinspace $E_{c.m.}$
reveals itself in so general setting, lends support to the idea that it can
survive notwithstanding model-dependent factors (electromagnetic radiation,
gravitational radiation, etc.). However, at present, this is only a
conjecture since, say, the role of gravitational radiation becomes more
significant when the particle's velocity approaches the speed of light \cite%
{berti}, \cite{ted}, \cite{force}. Also important is whether or not the
phenomenon is observable in more realistic astrophysical situations
including measurements which can be done at infinity \cite{flux}. These
issues deserve further study.

\end{document}